\documentclass[aps,prd,nopacs,onecolumn,nofootinbib,notitlepage,longbibliography]{revtex4-1}

\usepackage{amsfonts}
\usepackage{amsmath}
\usepackage{bm}
\usepackage{tipa}
\usepackage{longtable}

 \usepackage[colorlinks=true,hyperfootnotes=true,citecolor=cyan]{hyperref}

\newcommand{\be}{\begin{equation}}
\newcommand{\ee}{\end{equation}}
\newcommand{\ba}{\begin{eqnarray}}
\newcommand{\ea}{\end{eqnarray}}
\newcommand{\bs}{\begin{subequations}}
\newcommand{\es}{\end{subequations}}

\newcommand{\bbe}{\boldsymbol{\mathrm{e}}}
\newcommand{\e}{\mathrm{e}}
\newcommand{\ie}{\text{\textschwa}}
\newcommand{\bbie}{\boldsymbol{\textbf{\textschwa}}}

\newcommand{\bE}{\boldsymbol{E}}
\newcommand{\bJ}{\boldsymbol{J}}
\newcommand{\bL}{\boldsymbol{L}}
\newcommand{\bR}{\boldsymbol{R}}
\newcommand{\bQ}{\boldsymbol{Q}}
\newcommand{\bT}{\boldsymbol{T}}

\newcommand{\bba}{\boldsymbol{a}}

\newcommand{\bg}{\boldsymbol{g}}
\newcommand{\bh}{\boldsymbol{h}}
\newcommand{\bj}{\boldsymbol{j}}
\newcommand{\bl}{\boldsymbol{l}}

\newcommand{\bq}{\boldsymbol{q}}
\newcommand{\br}{\boldsymbol{r}}
\newcommand{\bss}{\boldsymbol{s}}
\newcommand{\bt}{\boldsymbol{t}}

\newcommand{\bGamma}{\boldsymbol{\Gamma}}
\newcommand{\bkappa}{\boldsymbol{\kappa}}
\newcommand{\blambda}{\boldsymbol{\lambda}}
\newcommand{\btheta}{\boldsymbol{\theta}}

\newcommand{\bdiff}{\boldsymbol{\mathrm{d}}}
\newcommand{\bDiff}{\boldsymbol{\mathrm{D}}}

\newcommand{\lp}{\left(}
\newcommand{\rp}{\right)}

\newcommand{\nn}{\nonumber}

\newcommand{\1}{1$^\text{st}$}
\newcommand{\2}{2$^\text{nd}$}
\newcommand{\3}{3$^\text{rd}$}

\begin{document}

\title{Noether charges in the geometrical trinity of gravity}

\author{Jose Beltr{\'a}n Jim{\'e}nez}\email{jose.beltran@usal.es}
\affiliation{Departamento de Física Fundamental and IUFFyM, Universidad de Salamanca, E-37008 Salamanca, Spain}

\author{Tomi S. Koivisto}\email{tomik@astro.uio.no}
\address{Laboratory of Theoretical Physics, Institute of Physics, University of Tartu, W. Ostwaldi 1, 50411 Tartu, Estonia}
\address{National Institute of Chemical Physics and Biophysics, R\"avala pst. 10, 10143 Tallinn, Estonia}

\begin{abstract}

The Noether currents are derived in a generic metric-affine theory of gravity, and the holographic nature of the gravitational entropy and energy-momentum is clarified. The main result is the verification of the canonical resolution to the energy-momentum problem in the Noether formalism. 

\end{abstract}

\maketitle

\section{Introduction}

Noether's 2 theorems \cite{Noether} were discovered in the context of seeking the answer to the problem of defining the energy of the gravitational field,
but the problem remains open today. The main difficulty resides in that general covariance, being a gauge symmetry, does not yield a non-vanishing Noether current that could make a good definition of energy. 
This problem can be fixed by resorting to some external geometry, and it is widely believed 
that gravitational energy can be unambiguously defined only with respect to some background structure, such as an asymptotically maximally symmetric
solution \cite{Misner1973}. The modern view is that gravitational energy can be defined quasi-locally \cite{Penrose:1982wp},  but it is fair to say that there is no physical 
principle that would distinguish amongst the various quasi-local prescriptions \cite{review,Chen:2015vya}.  

The canonical resolution was proposed recently \cite{BeltranJimenez:2019bnx,Koivisto:2019ggr}. The minimal ``universal background structure'' that may bestow covariance upon
otherwise pseudo-tensorial quantities is, tautologically, the pure-gauge translation connection. The implementation of this connection in Einstein's formulation
of General Relativity \cite{1916SPAW......1111E} results in the reformulation as an integrable gauge theory of translations \cite{Koivisto:2018aip} dubbed Coincident General Relativity \cite{BeltranJimenez:2017tkd}. The proposal was to identify the general-relativistic inertial frame, which fixes the additional gauge freedom due to the connection in this theory\footnote{Up to a global affine transformation. From a more field theoretical perspective, the integrable property of this reformulation can alternatively be interpreted in terms of St\"uckelberg fields that gauge the translations  \cite{Koivisto:2018aip,BeltranJimenez:2017tkd}, and the inertial frame is then understood as the one wherein the St\"uckelberg fields trivialise the energy-momentum tensor of the gravitational field.}, and thus uniquely determines gravitational energy, as the frame wherein the energy-momentum tensor of the gravitational field vanishes \cite{BeltranJimenez:2019bnx} (in agreement with Cooperstock's hypothesis \cite{Cooperstock1992-COOELI}). The proposal is consistent with the Noether's \2 theorem, and with the intuition that gravity is equivalent to inertia. 

However, the proposal is at odds with the usual considerations of energy in metric-teleparallel gravity \cite{moller,deAndrade:2000kr,Maluf:2002zc,Emtsova:2021ehh}. Though the gravitational energy-momentum introduced in the metric-teleparallel theory is generally covariant, it is not Lorentz covariant. Therefore calculations of this energy-momentum are as arbitrary as with Einstein's pseudo-tensor \cite{1916SPAW......1111E}, but as we will argue here, the criterion for the inertial frame \cite{BeltranJimenez:2019bnx} should yield uniquely the correct quasi-local energy also when applied in the metric-teleparallel theory. 

An important lesson of this letter will be that the gravitational energy-momentum and entropy are {\it purely holographic}, and this property is independent of the gravity theory and its formulation. Though the 
gravitational energy-momentum current $\bt_a$ vanishes in the inertial frame, we nevertheless find that gravitational waves can, and do carry energy (in contrast to Cooperstock's hypothesis \cite{Cooperstock1992-COOELI}). 
The reason is that the conserved charges arise from pure surface integrals.  The field equations of Coincident General Relativity  written in terms of the gravitational excitation $\bh_a$ and sourced by an energy-momentum $\bt_a$ have the form $\bDiff\bh_a = \bt_a$, and even though in vacuum $\bt_a =0$, there may exist non-vanishing conserved charges $\oint \bh_a$ enclosed within the surface of integration. 
 
Though this resolution had been already put forward \cite{BeltranJimenez:2019bnx,Koivisto:2019ggr}, its relation to the Noether theorems had not been clarified. 
This letter reports our finding that, indeed $\bj_a = \bh_a$ is the Wald current \cite{Wald:1993nt,Iyer:1994ys} such that $\bJ_a \approx \bdiff \bh_a$ is the Noether current associated with the translation invariance.
We seize the opportunity to extend the derivation to a fully general theory of gravity, since then we can adapt the results to the whole
``geometrical trinity of gravity'' \cite{Jimenez:2019woj}: see appendix \ref{interpretation} for a quick review of the triad of gravity's interpretations. 
In the next section \ref{currents} we obtain the form of the current in a generic metric-affine theory \cite{Hehl:1994ue,JimenezCano:2021rlu} and 
in the following section \ref{diffs} apply it to diffeomorphisms and analyse it in under the different teleparallelisms. Section \ref{conclusion} is the conclusion.  

\section{Currents in metric-affine gravity}
\label{currents}

We consider a Lagrangian $\bL$ that may depend upon the 3 independent fields, a metric $g^{ab}$, a GL-connection $\bGamma_a{}^b$ of the General Linear group, a coframe $\bbe^a$, and their derivatives 
through the non-metricity 1-form $\bQ^{ab}=-\bDiff g^{ab} = -\bdiff g^{ab} + 2\bGamma^{(ab)}$, the torsion 2-form $\bT^a = \bDiff \bbe^a = \bdiff \bbe^a - \bGamma_b{}^a\wedge\bbe^b$ and the curvature 2-form $\bR_a{}^b = \bdiff\bGamma_a{}^b + \bGamma_a{}^c\wedge\bGamma_c{}^b$. 
The variation of the Lagrangian is thus
\bs
\ba
\delta \bL & = & \delta g^{ab}\frac{\partial\bL}{\partial g^{ab}} - \delta\bbe^a\wedge \bt_a + \delta\bQ^{ab}\wedge\bq_{ab} 
 + \delta\bT^a\wedge\bss_a + \delta\bR_a{}^b\wedge\br^a{}_b \label{implicit} \\
 & = & \delta g^{ab}\bg_{ab} + \delta\bbe^a\wedge\bE_a + \delta\bGamma_a{}^b\wedge\bE^a{}_b 
  +  \bdiff\lp -\delta g^{ab}\bq_{ab} + \delta\bbe^a\wedge\bss_a + \delta\bGamma_a{}^b\wedge\br^a{}_b\rp\,, \label{explicit}
\ea
\es
where the conjugates can be understood to be implicitly defined by (\ref{implicit}),
\bs
\be
\bq_{ab}   =  \frac{\partial\bL}{\partial \bQ^{ab}}\,, \quad
\bt_a = -\frac{\partial\bL}{\partial \bbe^{a}}\,, \quad 
\bss_{a}   =  \frac{\partial\bL}{\partial \bT^{a}}\,, \quad
\br^a{}_b   =   \frac{\partial\bL}{\partial \bR_a{}^b}\,, 
\ee
and the EoM's in (\ref{explicit}) are
\be
\bg_{ab}  =  \frac{\partial\bL}{\partial g^{ab}} + \bDiff\bq_{ab}\,, \quad
\bE_a  =   \bDiff\bss_a - \bt_a\,, \quad
\bE^a{}_b  =  2\bq^a{}_b - \bbe^a\wedge\bss_b + \bDiff\br^a{}_b\,. 
\ee
\es
The on-shell variation of the Lagrangian is therefore given by the following symplectic current 3-form:
\bs
\be \label{symplectic}
\btheta = -\delta g^{ab}\bq_{ab} + \delta\bbe^a\wedge\bss_a + \delta\bGamma_a{}^b\wedge\br^a{}_b\,.
\ee
By considering a symmetry of the system corresponding to a transformation of the fields that shifts the Lagrangian by a total derivative $\delta \bL = \bdiff \bl$ with $\bl$ some 3-form, we obtain the Noether current as
\be \label{noether}
\bJ = \btheta - \bl\,,
\ee
\es
which, by construction, is conserved on-shell $\bdiff\bJ \approx 0$. 

As the \1 example, let us consider the GL transformation by a parameter $\alpha_a{}^b$, so that $\delta g^{ab} = \alpha_c{}^a g^{cb} +  \alpha_c{}^b g^{ac}$, 
$\delta \bbe^a = \alpha_b{}^a\bbe^b$ and $\delta\bGamma_a{}^b = -\bDiff\alpha_a{}^b$. 
Since $\bL$ is GL-invariant, the shift 3-form is trivial $\bl=0$. {We can then use the general variation in \eqref{explicit} to obtain
\ba
\delta \bL  &=& -2\alpha^{(ab)}\bg_{ab} + \alpha_b{}^a\bbe^b\wedge\bE_a - \bDiff\alpha_a{}^b\wedge\bE^a{}_b 
  +  \bdiff\lp -2\alpha^{(ab)}\bq_{ab} + \alpha_b{}^a\bbe^b\wedge\bss_a - \bDiff\alpha_a{}^b\wedge\br^a{}_b\rp\nonumber\\
&=&  \alpha_a{}^b\Big( -2g^{ac}\bg_{cb} + \bbe^a\wedge\bE_b + \bDiff\bE^a{}_b \Big)
  +  \bdiff\lp -2\alpha^{(ab)}\bq_{ab} + \alpha_b{}^a\bbe^b\wedge\bss_a - \bDiff\alpha_a{}^b\wedge\br^a{}_b-\alpha_a{}^b\bE^a{}_b\rp\,.
\ea
From this expression we can obtain the usual Bianchi identities by choosing a vanishing gauge parameter at infinity so that the boundary term drops,
\be
g^{ac}\bg_{cb} + \bbe^a\wedge\bE_b + \bDiff\bE^a{}_b=0\,.
\ee
We are however interested in the physical on-shell conserved current associated to global symmetries that is obtained from the boundary term that can be written in the very simple form
\be\label{alphacurrent0}
 -2\alpha^{(ab)}\bq_{ab} + \alpha_b{}^a\bbe^b\wedge\bss_a - \bDiff\alpha_a{}^b\wedge\br^a{}_b-\alpha_a{}^b\bE^a{}_b= -\bdiff\lp\alpha_a{}^b\br^a{}_b\rp.
 \ee
This is of course the same on-shell conserved current obtained from \eqref{noether} as
\ba \label{alphacurrent}
\bJ & = & \btheta = -2\alpha^{(ab)}\bq_{ab} + \alpha_b{}^a\bbe^b\wedge\bss_a - \bDiff\alpha_a{}^b\wedge\br^a{}_b = -\bdiff\lp\alpha_a{}^b\br^a{}_b\rp + \alpha_a{}^b\bE^a{}_b\,. 
\ea
that coincides with \eqref{alphacurrent0} up to an irrelevant term proportional to the equations of motion that vanishes on-shell. This simple calculation demonstrates the property of gauge transformations that the on-shell conserved current} admits a potential 2-form $\bj$ so that $\bJ \approx \bdiff \bj$ and the charge turns into a pure boundary term. In the following we call the exact form $\bj$ the Wald current, due to its role, in the case of diffeomorphisms, as the determinant of the black hole entropy \cite{Wald:1993nt}. Wald and Iyer \cite{Iyer:1994ys} have derived 2 equivalent forms of the current. Incidentally, choosing a radial boost $\alpha^{ab} = \alpha^{[tr]}$ of the horizon, the expression (\ref{alphacurrent}) gives the alternative form. It is our \1 new insight that the entropy of a black hole can also be understood as a Lorentz charge.

\section{Diffeomorphism charges}
\label{diffs}

Consider the diffeomorphism transformation $\delta = \mathcal{L}_\xi$ along the vector $\xi^\mu$. It is given by the ``Cartan's magic formula''
\bs
\be  \label{lie1}
\mathcal{L}_\xi \bba = \xi\lrcorner(\bdiff\bba) + \bdiff(\xi\lrcorner\bba)\,,
\ee
for an arbitrary $p$-form $\bba$. There are various alternatives \cite{Obukhov:2006ge,Prabhu:2015vua,Elgood:2020svt}, such as the covariant diffeomorphism \cite{Hehl:1994ue}
\be \label{lie2}
\tilde{\mathcal{L}}_\xi \bba = \xi\lrcorner(\bDiff\bba) + \bDiff(\xi\lrcorner\bba)\,. 
\ee
\es
It follows  \cite{Obukhov:2006ge} that the consistent action on the connection is $\tilde{\mathcal{L}}_\xi \bGamma_a{}^b = \xi\lrcorner \bR_a{}^b$. 
Since $\bL$ is a scalar spacetime volume form, under either $\mathcal{L}_\xi$ or $\tilde{\mathcal{L}}_\xi$ one can show (see e.g. \cite{Hehl:1994ue}) 
\be
{\bl} = \xi\lrcorner\bL =  -\lp\xi\lrcorner\bbe^a\rp\bt_a + \lp\xi\lrcorner\bQ^{ab}\rp\bq_{ab} 
 + \lp\xi\lrcorner\bT^a\rp\wedge\bss_a + \lp\xi\lrcorner\bR_a{}^b\rp\wedge\br^a{}_b\,.  
\ee
Next we apply (\ref{lie2}) to the symplectic current (\ref{symplectic}) and perform the partial integrations,
\bs
\label{covariantcurrents}
\be \label{apply2}
\tilde{\btheta} = \bdiff\tilde{\bj}
+ \lp\xi\lrcorner \bQ^{ab}\rp\bq_{ab} + \lp\xi\lrcorner\bT^a\rp\wedge\bss_a - \lp\xi\lrcorner\bbe^a\rp\wedge\bDiff\bss_a + \lp\xi\lrcorner\bR_a{}^b\rp\wedge\br^a{}_b\,,
\ee
where the Wald current 2-form is
\be \label{wald}
\tilde{\bj} =  \lp\xi\lrcorner\bbe^a\rp\bss_a\,,
\ee
and the Noether current as defined in (\ref{noether}) is
\be
\tilde{\bJ} =  \bdiff\tilde{\bj} - \lp\xi\lrcorner\bbe^a\rp\bE_a\,.
\ee
\es
The bulk piece vanishes on-shell, as expected. 
Applying instead (\ref{lie1}) to the symplectic current, it is useful to note that
\be
\mathcal{L}_\xi \bGamma_a{}^b = \bDiff\lp \xi\lrcorner\bGamma_a{}^b\rp + \xi\lrcorner\bR_a{}^b\,,
\ee
where the $\bDiff$ acts on $\xi\lrcorner\bGamma_a{}^b$ as it would upon a tensor. Now we get instead of (\ref{covariantcurrents}), 
\bs
\label{noncovariantcurrents}
\ba \label{apply1}
\btheta & = & \bdiff\bj
- \lp\xi\lrcorner \bdiff g^{ab}\rp\bq_{ab} + \lp\xi\lrcorner\bdiff\bbe^a\rp\wedge\bss_a - \lp\xi\lrcorner\bbe^a\rp\bdiff\bss_a + \lp\xi\lrcorner\bR_a{}^b\rp \wedge\br^a{}_b - \lp\xi\lrcorner\bGamma_a{}^b\rp\bDiff\br^a{}_b \nn \\
& = &  \bdiff\bj + \lp\xi\lrcorner \bQ^{ab}\rp\bq_{ab}  + \lp\xi\lrcorner\bT^a\rp\wedge\bss_a  -  \lp\xi\lrcorner\bbe^a\rp\bDiff\bss_a 
+ \lp\xi\lrcorner\bR_a{}^b\rp \wedge\br^a{}_b - \lp\xi\lrcorner\bGamma_a{}^b\rp\bE^a{}_b\,,
\label{symplectic2} \\
{\bj} & = &  \lp\xi\lrcorner\bbe^a\rp\bss_a + \lp\xi\lrcorner\bGamma_a{}^b\rp\br^a{}_b\,,  \label{wald2} \\
\bJ & = & \bdiff\bj - \lp\xi\lrcorner\bbe^a\rp\bE_a - \lp\xi\lrcorner\bGamma_a{}^b\rp\bE^a{}_b\,,  \label{noether2} 
\ea
\es
for the symplectic, the Wald, and the Noether currents, respectively. (For the Noether identities from (\ref{lie1}), see Lemma 4 of \cite{BeltranJimenez:2020sih}, and for the Noether identities from (\ref{lie2}), see the section 5.2.1 of 
\cite{Hehl:1994ue}). After deriving the general expressions (\ref{covariantcurrents}) and (\ref{noncovariantcurrents}), we shall apply it to each of the cases in the ``geometrical trinity'' \cite{Jimenez:2019woj}.

\subsection{Einstein-Cartan}

Only the curvature features in the well-known Einstein-Cartan action. Thus we can set $\bss_a=0$ in (\ref{wald}), and the current disappears. The gauge symmetry is trivial. We may resort to the non-covariant alternative (\ref{wald2}). However, there the on-shell $\bGamma_a{}^b$ is determined to be the Levi-Civita connection only up to a projective transformation. Another source of ambiguity is of course that  $\bGamma_a{}^b$ is not a tensor, and therefore the $\bj$ depends arbitrarily upon the reference frame. 

\subsection{Teleparallelism}

The teleparallel equivalent of General Relativity without the symmetric nor the metric constraint was introduced in Ref.\cite{BeltranJimenez:2019odq}.
The flat geometry is imposed by $\bL \rightarrow \bL + \blambda^a{}_b\wedge\bR_a{}^b$. 
Then we only have a new EoM $\bR_a{}^b \approx 0$. The currents are unaffected, one only has to take into account that now $\br^a{}_b = \blambda^a{}_b$. 
The Bianchi identity $\bDiff\bE^a{}_b =0$ gives
\be
2\bDiff\bq^a{}_b - \bT^a\wedge\bss_b + \bbe^a\wedge\bDiff\bss_b \approx 0\,. 
\ee
Using the other EoM we can also write $2\bDiff\bq^a{}_b - \bT^a\wedge\bss_b + \bbe^a\wedge\bt_b \approx 0$. 
These equations determine the dynamics, but they do not completely determine\footnote{The underlying reason is that the Lagrange multipliers possess gauge symmetries \cite{BeltranJimenez:2018vdo}.} the Lagrange multiplier $\blambda^a{}_b$ which is decoupled from the dynamics. 
We only obtain
\be
\bDiff\blambda^a{}_b \approx 2\bq^a{}_b - \bbe^a{}\wedge\bss_b\,. 
\ee
The Wald current 2-form (\ref{wald2}) is therefore, in general, undetermined. We can, however, compute the conserved charge as a surface integral by choosing a gauge wherein 
$\xi\lrcorner\bGamma_a{}^b \overset{S}{=} 0$, or as a volume integral in a gauge wherein $\bDiff (\xi\lrcorner\bGamma_a{}^b) \overset{V}{=} 0$.
Obviously, the gauge $\bGamma^a{}_b=0$ eliminates the dependence of the current upon the unknown Lagrange multiplier. We then find that the conserved charges are given by the surface integral $\oint \xi\lrcorner\bbe^a\bss_a$. 
The covariant current (\ref{wald}) yields consistently this same result, without the need to fix the ``generalised Weitzenb\"ock'' $\bGamma^a{}_b=0$ or any other gauge.  

\subsection{Symmetric-teleparallelism}

Let us now consider that $\bL \rightarrow \bL + \blambda_a\wedge\bT^a + \blambda^a{}_b\wedge\bR_a{}^b$.
We have then $\br^a{}_b = \blambda^a{}_b$ and $\bss_a=\blambda_a$, and the EoM's are
\be
\bE_a  =  \bDiff\blambda_a - \bt_a\,, \quad \bE^a{}_b  =  2\bq^a{}_b - \bbe^a\wedge\blambda_b + \bDiff\blambda^a{}_b\,.
\ee
From $\bDiff\bE^a{}_b=0$ we get now $2\bDiff\bq^a{}_b + \bbe^a\wedge\bDiff\blambda_b \approx 0$, and we can invert this to solve for 
\be
\bDiff\blambda_a \approx -2\bbie_b\lrcorner \bDiff\bq^b{}_a\,. 
\ee
This is nothing but the ``remarkable relation'' (18) of Ref.\cite{Koivisto:2019ggr}. It was shown to have the unique solution\footnote{Again, the Lagrange multiplier is subject to a gauge symmetry that only permits to obtain it up to the covariant exterior derivative of some arbitrary 1-form. This redundancy is eliminated by the requirement of local, linear and parity-invariant constitutive law \cite{Koivisto:2019ggr}.} $\blambda_a = \bh_a$, where $\bh_a$ is the gravitational
excitation 2-form (and $\bq_{ab}$ is the premetric mass excitation 3-form, the components $(\star q_{ab})^\alpha\e^a{}_\mu\e^b{}_\nu = P^\alpha{}_{\mu\nu}$ being a.k.a. the non-metricity conjugate tensor). The explicit form of the gravitational excitation is (with $m_P$ the Planck mass and $\epsilon_{abcd}$ the totally antisymmetric Levi-Civita symbol) 
\be
\bh_a = \frac{m_P^2}{2}\epsilon_{ab}{}^{cd}\lp \bbie_c\lrcorner\bQ_{de}\rp\bbe^b\wedge\bbe^e\,. 
\ee
In the coincident gauge, the components of $\bh_a$ reduce to the von Freud superpotential \cite{10.2307/1968929}. 
The dynamics of the theory are thus given by
\be
\bDiff\bh_a  \approx \bt_a\,, 
\ee
since the coframe EoM $\bDiff \bt^a \approx 0$ is an identity. The conserved charges corresponding to a translation by $\xi^a = \xi\lrcorner\bbe^a$ are therefore given 
by the surface integral $\oint  \xi^a \bh_a$. This is the gauge-invariant and coordinate-invariant expression used (in its tensorial form) in Refs.\cite{BeltranJimenez:2019bnx,Koivisto:2019ggr}.  

\subsection{Metric-teleparallelism}

Finally, we also look at the special case $\bL \rightarrow \bL +  \blambda^a{}_b\wedge\bR_a{}^b + \bkappa_{ab}\wedge\bQ^{ab}$.
The dynamics can now be determined from 
\be
\bDiff\bss^a \approx  \bt^a \,, \quad \bDiff\bt^a \approx 0\,.
\ee
Now we cannot determine $\blambda_a{}^b$ but could determine $\bDiff\bkappa_{ab}$. With the Wald current, the situation is similar as without the metric restriction. We may argue that the inertial frame is correctly determined as $\bt^a =0$ also in metric-teleparallel equivalent of General Relativity, since in the $\bGamma_a{}^b=0$ gauge the superpotential $\bss_a$ should coincide with the gravitational excitation $\bh_a$ in the symmetric-teleparallel formulation.  

\section{Conclusion}
\label{conclusion}

In this letter we have obtained the Noether currents in generic metric-affine theory of gravitation and we have emphasised the holographic nature of the gravitational entropy and of energy-momentum. These are computed
as volume integrals over Noether currents that reduce to surface integrals over Wald currents. 

We \1 considered the GL currents and found that the Wald entropy could equivalently be understood as a Lorentz charge. Then, we considered the diffeomorphism currents, and established that the ambiguities in the quasi-local energy-momentum can be eliminated under teleparallelism, with the physical requirement that $\bt_a$ may only be nonzero due to matter. The derivations are technically neater for the covariant version (\ref{lie2}), but the main conclusion is the same for (\ref{lie1}). The claim is that the energy-momentum current $\bt_a$ (which is usually considered in metric-teleparallel gravity), is only generated by non-inertial effects, but the gravitational energy-momentum $\oint\bj$ is revealed at the boundary. This determines uniquely, for example, the physical characteristics of gravitational waves. 

It is amusing to point out that the standard metric- and the symmetric-teleparallel pictures \cite{Jimenez:2019woj} are mirror images of each other in the way that the Weitzenb\"ock connection 
$\Gamma^\alpha{}_{\mu\nu} = \e_a{}^\alpha\partial_\mu\ie^a{}_\nu$ corresponds to the trivial gauge field $\bGamma_a{}^b=0$ whilst the coincident gauge
$\Gamma^\alpha{}_{\mu\nu} = 0$ may entail non-trivial gauge geometry $\bGamma_a{}^{b} =(\bdiff\bbie_a{})\lrcorner  \bbe^b$. 
The interpretation in the former case is that despite trivial gauge geometry there are forces (torsion) distorting the spacetime, whilst in the latter case 
spacetime is a priori integrable but there are underlying inertial interactions (nonmetricity). Our findings for the Noether charges in this letter are equivalent for those 2 very different interpretations. The minimal matter coupling is one principle that can distinguish the ``physical'' geometry \cite{BeltranJimenez:2020sih}.

The main result was to confirm, in the framework of Noether formalism, the canonical resolution to the problem of localising the gravitational energy. The resolution had been arrived at  \cite{BeltranJimenez:2019bnx} by following Einstein's original physical reasoning  \cite{1916SPAW......1111E} and it was formally deduced from \1 principles assuming nothing but the inevitable axioms of the premetric program \cite{Koivisto:2019ggr}. The same simple result is reached by the canonical Noether procedure. 

\begin{acknowledgments}
We would like to thank the anonymous referee for valuable questions, recommendations and comments. This work was supported by the Estonian Research Council grants PRG356 “Gauge Gravity” and MOBTT86, and by the European Regional Development Fund CoE program TK133 “The Dark Side of the Universe”. J.B.J. acknowledges support from the Atracci\'on del Talento Cient\'ifico en Salamanca program and from the project PGC2018-096038-B-I00 by Spanish Ministerio de Ciencia, Innovaci\'on y Universidades.
\end{acknowledgments}

\appendix

\section{Interpretations of gravity in the geometrical trinity}
\label{interpretation}

To elucidate the physical interpretations of gravity, it is better to use the tensor formalism, since physics does not take place in a tangent space. We shall consider the response of matter 
to gravity, since that is independent of the particular theory of gravity and only depends on the matter coupling. In the standard formulations of teleparallel gravity, the matter coupling is effectively the metric one, and thus the trajectory of a test particle is given by the geodesic equation (in symmetric-teleparallism it is the consequence of the minimal coupling gauge principle, whereas in metric-teleparallelism it is an extra assumption that has to be postulated for the consistency of the theory \cite{BeltranJimenez:2020sih}).

Thus, let us commence by recalling the geodesic equation that is given in terms of the Levi-Civita connection of the metric as follows:
\be \label{rg}
\ddot{x}^\alpha + \overbrace{\left\{^{\phantom{i} \alpha}_{\mu\nu}\right\}}^{\textcolor{red}{\text{geometry}}}\dot{x}^\mu\dot{x}^\nu = 0\,. \nn
\ee
The standard interpretation of this equation is that particles move along straight lines in a curved manifold and, hence, it has a geometrical nature.

In metric-teleparallelism, the {\it same} equation can be written, using the identity
 $\left\{^{\phantom{i} \alpha}_{\mu\nu}\right\}\dot{x}^\mu\dot{x}^\nu=
 (\Gamma^\alpha{}_{\mu\nu}-\frac{1}{2}T_{\mu\nu}{}^\alpha{})\dot{x}^\mu\dot{x}^\nu$ 
 with $\Gamma^\alpha{}_{\mu\nu}$ and $T^\alpha{}_{\mu\nu}=2\Gamma^\alpha{}_{[\mu\nu]}$ the Weitzenb\"ock connection and its torsion respectively, as 
\be \label{mtgeod}
\ddot{x}^\alpha +  \overbrace{{\Gamma}^\alpha{}_{\mu\nu}}^{\textcolor{red}{\text{Weitzenb\"ock}}}\dot{x}^\mu\dot{x}^\nu = \overbrace{-\frac{1}{2}T_{\mu\nu}{}^\alpha{}\dot{x}^\mu\dot{x}^\nu}^{\textcolor{red}{\text{force }\ F^\alpha}} \,. \nn
\ee
The Weitzenb\"ock connection cannot be globally eliminated. The interpretation of the RHS as a force is appropriate since $F^\alpha$ is orthogonal to the 4-velocity $\dot{x}^\alpha$. 

Yet a \3 equivalent form of the same equation is found in symmetric-teleparallelism by using the relation between the symmetric-teleparallel and Levi-Civita connections given in terms of the non-metricity 
$Q_\alpha{}^{\mu\nu} = -\nabla_\alpha g^{\mu\nu}$, 
\be \label{stgeod}
\ddot{x}^\alpha + \overbrace{\Gamma^\alpha{}_{\mu\nu}}^{\textcolor{red}{\text{pure gauge}}}\dot{x}^\mu\dot{x}^\nu =  \overbrace{\lp\frac{1}{2}Q^\alpha{}_{\mu\nu}-Q_{\mu\nu}{}^\alpha\rp\dot{x}^\mu\dot{x}^\nu}^{\textcolor{red}{\text{inertia}\ I^\alpha}}\,. \nn
\ee
The pure-gauge connection can be eliminated globally in the unitary, or the  ``coincident'' gauge $\Gamma^\alpha{}_{\mu\nu} = 0$. The RHS does not admit an interpretation as a force, since in general we have
$I^\alpha\dot{x}_{\alpha} \neq 0$. 

From this brief discourse we can conclude that gravity admits equivalent interpretations either in terms of geometry, force or inertia.

\bibliography{Q2021F}

\end{document}